\documentclass[epj]{svjour} 

\usepackage{graphics}

\usepackage{epsfig}

\makeindex

\def\text{{}}   \newcommand{\tr}{{\rm Tr}}
\def\lsim{\raise0.3ex\hbox{$<$\kern-0.75em\raise-1.1ex\hbox{$\sim$}}}
\def\gsim{\raise0.3ex\hbox{$>$\kern-0.75em\raise-1.1ex\hbox{$\sim$}}}

\begin{document}

\title{ Static quark anti-quark free and internal energy in 2-flavor QCD }

\author{Olaf Kaczmarek\inst{1} and Felix Zantow\inst{2}} \institute{
  Fakult\"{a}t f\"{u}r Physik, Universit\"{a}t Bielefeld, D-33615 Bielefeld,
  Germany \and Physics Department, Brookhaven National Laboratory, Upton, NY
  11973 USA } \date{\today}

\abstract{We study the change in free and internal energy due to the presence
  of a heavy quark anti-quark pair in a thermal heat bath in QCD with 2-flavors
  of staggered quarks at finite temperature. We discuss string breaking below
  as well as screening above the transition. Similarities and differences to
  the quenched case are discussed.  }


\maketitle

\section{Introduction}
The study of the fundamental forces between quarks is an essential key to the
understanding of QCD and the occurrence of different phases at high
temperatures ($T$) and/or non-zero quark chemical potential ($\mu$). The free
energy of a static quark anti-quark pair \cite{McLerran:1981pb}, separated by
distance $r$, is an important tool to analyze the $r$ and $T$ dependence of the
QCD forces. Similar to the free energies also the internal energies have been
introduced \cite{Kaczmarek:2002mc,Zantow:2003ui} and are expected to play an
important role in the discussion of quarkonia binding properties
\cite{Brown:2004qi,Wong:2004kn,Shuryak:2004tx}. The structure of these
observables in the short and intermediate distance regime, $rT\;\lsim\;1$, is
relevant for the discussion of in-medium modifications of heavy quark bound
states
\cite{Brambilla:2004wf,Digal:2001iu,Digal:2001ue,Wong:2001kn,Wong:2001uu} which
are sensitive to thermal modifications of the heavy quark potential
\cite{Matsui:1986dk}. Up to quite recently
\cite{Kaczmarek:2003ph,Petreczky:2004pz}, however, most of these discussions
concerned quenched QCD.
 
Several qualitative differences, however, are to be expected when taking the
influence of dynamical fermions into account; the phase transition in QCD will
appear as an crossover rather than a 'true' phase transition with related
singularities in thermodynamic observables. Moreover, in contrast to a steadily
increasing confinement interaction in quenched QCD, in full QCD the strong
interaction below $T_c$ will show a qualitative different behavior due to the
possibility of string breaking. Thus it is quite important to extend our
developed concepts for the analysis of free energies and internal energies in
quenched QCD \cite{Kaczmarek:2002mc,Kaczmarek:2003dp,Kaczmarek:2004gv} to the
complex case of QCD. This has recently been done for $2$- and $3$-flavor QCD
\cite{Kaczmarek:2003ph,okacz05/1,okacz05/2}.

\section{Quark anti-quark free energy}
We will discuss here lattice results for the quark anti-quark free and internal
energies in $2$-flavor QCD ($N_f=2$) using an improved staggered fermion action
with quark mass $m/T=0.4$. Any further detail on this study can be found in
\cite{Kaczmarek:2003ph,okacz05/1,okacz05/2}. For information on the improved actions
used in these simulations see \cite{Allton:2002zi,Allton:2003vx,Karsch:2000kv}.
While in earlier studies of the heavy quark free energy in quenched
\cite{Kaczmarek:1999mm} and full QCD \cite{DeTar:1998qa} the color averaged
operators were analyzed we discuss here the free and also the internal energies
in the color singlet channel. The static quark sources are described by the
Polyakov loop, $W(\vec{x}) = \prod_{\tau=1}^{N_\tau} U_0(\vec{x},\tau)$ with
$U_0(\vec{x},\tau) \in SU(3)$ being defined on the lattice link in time
direction.  The free energies in the color singlet channel is then given by
\cite{McLerran:1981pb,Nadkarni:1986cz}
\begin{eqnarray}
e^{-F_1(r)/T+C}&=&\frac{1}{3} \tr \langle  W(\vec{x}) W^{\dagger}(0) \rangle 
\label{f1}\;,
\end{eqnarray}
where $r=|\vec{x}|$ and $C$ is a suitably chosen renormalization constant. The
operator used here to calculate $F_1(r,T)$ is not gauge invariant. Our
calculations thus have been performed in Coulomb gauge
\cite{Philipsen:2002az}~.

\subsection{Renormalized free energy}
\begin{figure}[t]
  \epsfig{file=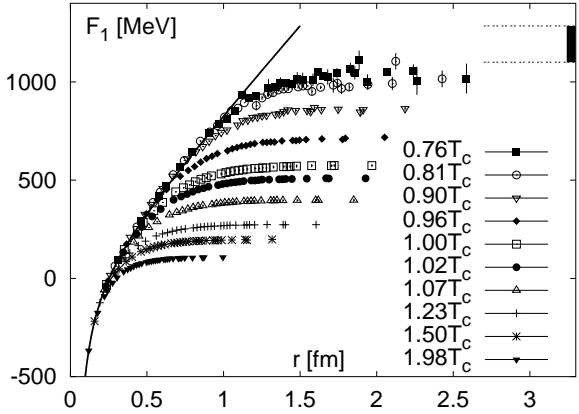,width=8.5cm}
\caption{
  Results for the renormalized color singlet free energies calculated in
  $2$-flavor QCD for $0.75 < T < 2.00$. The solid line represents the heavy
  quark potential at zero temperature, $V(r)$, given in \cite{zantow05}. At
  $T=0$ string breaking is expected at distances about $1.3\;-\;1.5$ fm
  \cite{Pennanen:2000yk}. The corresponding energies,
  $V(1.3\;-1.5\;\mbox{fm})\simeq1100\;-\;1285$ MeV, are shown by the band.  }
\label{plcsf}
\end{figure}
As noted above, the quark anti-quark free energy (\ref{f1}) and the expectation
value of the Polyakov loop suffer from divergences and require renormalization.
We follow here the conceptual approach recently suggested for the quark
anti-quark free energies in quenched QCD \cite{Kaczmarek:2002mc} and apply it
to our lattice results in full QCD. First experiences with this renormalization
prescription in QCD are reported in \cite{Kaczmarek:2003ph,Petreczky:2004pz}.
Renormalization prescriptions for the Polyakov loop expectation value are
discussed in \cite{Dumitru:2003hp,Zantow:2003uh}.

At distances much smaller than the inverse temperature, $r\ll 1/T$, the
dominant scale is set by $r$ and the running coupling will be controlled by
this scale, $r \ll 1/\Lambda_{QCD}$ \cite{zantow05,Kaczmarek:2004gv}. In this
limit the free energies are dominated by one-gluon exchange, {\em i.e.}  are
given by the heavy quark potential,
\begin{eqnarray}
F_1(r,T)&\equiv&V(r)\;\simeq\;-\frac{4}{3}\frac{\alpha(r)}{r}\;.\label{smallr}
\end{eqnarray}
We have neglected here any constant contributions to the free energy which, in
particular, will dominate the large distance behavior of the free energy.
Moreover, we already anticipated here the running of the coupling with the
dominant scale $r$. Following \cite{Kaczmarek:2002mc} we can use this property
to fix the constant $C$ in (\ref{f1}) by matching $F_1(r,T)$ to the zero
temperature heavy quark potential at small distance, $F_1(r\ll1/T,T)\simeq
V(r)$. Once the free energy is fixed at small distances also the large distance
behavior is fixed as no additional divergences get introduced at finite
temperature.

Our lattice results for $F_1(r,T)$ calculated in $2$-flavor QCD are summarized
in Fig.~\ref{plcsf} at several temperatures in the vicinity of the phase
transition. Due to renormalization the free energies coincide with the heavy
quark potential, $V(r)$, at small distances ($V(r)$ is specified in
\cite{okacz05/1,okacz05/2,zantow05}). When going to larger distances, however, thermal
modifications become important and the different effects from color screening
($T\;\gsim\;T_c$) and string breaking ($T\;\lsim\;T_c$) can be studied.

\subsection{Color screening and string breaking}\label{secshort}
\begin{figure}[t]
  \epsfig{file=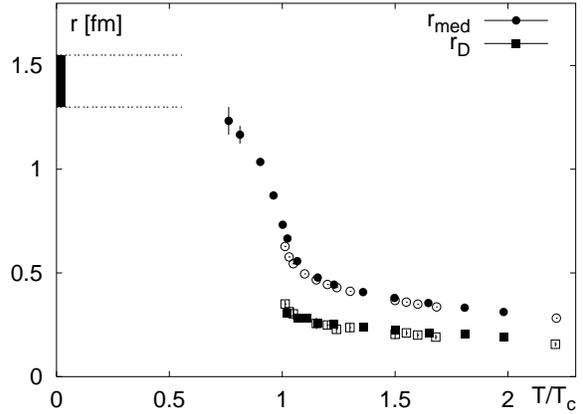,width=8.5cm}
\caption{The Debye screening radius, $r_D(T)$, and the scale $r_{med}(T)$ as
  function of temperature in physical 
units obtained from $2$-flavor QCD lattice studies 
(filled symbols). 
The open symbols correspond to calculations in 
quenched QCD \cite{Kaczmarek:2002mc,Kaczmarek:2004gv}. 
Further analysis of these data with respect to quarkonium 
dissociation temperatures can be found in \cite{Karsch}.}
\label{rmed}
\end{figure}
As the free energies shown in Fig.~\ref{plcsf} rapidly change from the zero
temperature like behavior at small distances to an almost constant behavior at
large distances we introduce here scales that may characterize this qualitative
change. For this purpose we introduced in \cite{Kaczmarek:2002mc} a scale
$r_{med}$ defined through $F_\infty(T)=V(r_{med})$, where $F_\infty(T)$ is the
plateau value which the free energy approaches at large distances,
$F_\infty(T)\equiv \lim_{r\to\infty}$ $F_1(r,T)$. This scale characterizes the
typical distance at which string breaking and color screening become relevant.
Alternatively one can at high temperature characterize screening properties of
the QCD plasma ($T\;\gsim\;T_c$) in terms of the Debye screening length,
$r_D(T)$, which is commonly defined as the inverse of the screening mass,
$r_D(T)\equiv1/m_D(T)$. We also estimated this scale by extracting the
screening masses non-perturbatively from the exponential fall off that the free
energies show above $T_c$ at large distances. Our results for
$r_D(T\;\gsim\;T_c)$ and $r_{med}$ are shown in Fig.~\ref{rmed} as function of
$T/T_c$ and are compared to findings in quenched QCD
\cite{Kaczmarek:2002mc,Kaczmarek:2004gv}. In general we find that both scales
decrease with increasing temperature and $r_{med}(T)\;\gsim\;r_D(T)$. We note
that both scales are expected to behave like $1/gT$ in the high temperature
limit. On the other hand, when comparing both scales to the findings in
quenched QCD no or only little differences in the temperature range
$1.2\;\lsim\;T/T_c$ $\lsim\;2$ could be identified. However, at temperatures
close to and below the transition, $T\;\lsim\;1.2T_c$, differences become quite
apparent. These differences signal the qualitative change in the phase
transition when changing from quenched ($1^{st}$-order transition) to full QCD
(crossover). In QCD $r_{med}$ is finite also below $T_c$
due to string breaking and shows a rapid increase with decreasing
temperatures\footnote{In quenched QCD $F_1(r,T<T_c)$ signals strict
confinement and $r_{med}$ is infinite.}.
In fact, the distance where the string is commonly expected to break at $T=0$
\cite{Pennanen:2000yk}, $r_{med}(T=0)\simeq 1.3\;-\;1.5$ fm, is almost
approached already at $T\simeq0.8T_c$. The quark anti-quark free energies in
$2$-flavor QCD thus are expected to show only little deviations from $V(r)$ at
$T=0$ at $T\;\gsim\;0.8T_c$.

\begin{figure}[tbp]
  \epsfig{file=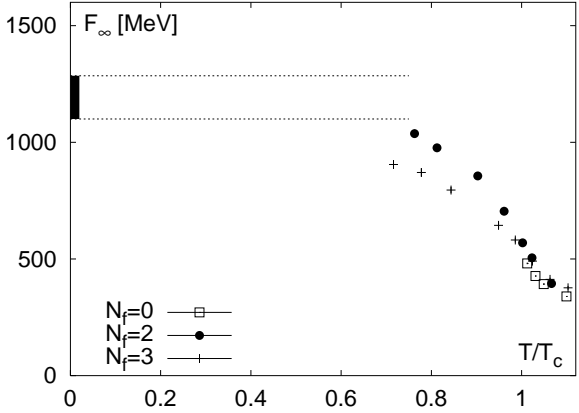,width=8.5cm}
\caption{The plateau value of the quark anti-quark free energy, $F_\infty(T)$, 
calculated in $2$-flavor QCD as function of $T/T_c$ at temperatures in the
vicinity 
and below the phase transition. The dashed band show again the string breaking 
energies at $T=0$, $V(r_{\text{break}})$, with 
$r_{\text{break}}\simeq1.3\;-\;1.5$ fm \cite{Pennanen:2000yk}. 
The open symbols ($T\;\gsim\;T_c$) correspond to $F_\infty(T)$ in quenched QCD 
(from Ref.~\cite{Kaczmarek:2002mc}) and the crosses to $3$-flavor QCD studies 
(from \cite{Petreczky:2004pz}).}
\label{string1}
\end{figure}
This is also evident when analyzing the plateau values, $F_\infty(T)$,
summarized in Fig.~\ref{string1} as function of $T/T_c$ and again
compared to recent findings in quenched
\cite{Zantow:2003ui} and $3$-flavor QCD \cite{Petreczky:2004pz} for
temperatures in the vicinity and below $T_c$.\footnote{The comparison of
  $F_\infty(T)$ performed here clearly
  depends on the relative normalization of the corresponding heavy quark
  potentials used for renormalization. We fixed these over all contributions
  such that the Cornell parameterization of $V(r)$ in all cases ($N_f=0,2,3$)
  contains no constant contribution.}
At $T\simeq0.8T_c$ $F_\infty(T)$
still is compatibel with the $T=0$ values, $V(r=\infty)\simeq 1.1$ GeV given in
\cite{Digal:2001iu,Pennanen:2000yk}. It, however, rapidly drops in the vicinity
of the transition to about half of this value, $F_\infty(T_c)\simeq 570$ MeV.
Although some flavor dependence can clearly be identified when comparing
results in $2$- and $3$-flavor QCD the value at $T_c$ is almost the same. A
similar value is also found in quenched QCD just above $T_c$. At very high
temperatures perturbation theory \cite{Gava:1981qd} suggests also negative
values for $F_\infty(T)$, {\em i.e.}
\begin{eqnarray}
F_\infty(T)&\simeq&-\frac{4}{3}m_D(T)\alpha(T)\;\simeq\;-{\cal O}(g^3T)\;.\label{PT}
\end{eqnarray}
Such a behavior has indeed been observed in lattice studies of quenched QCD at
high temperatures \cite{Kaczmarek:2002mc,Zantow:2003ui,Kacze05b}.

\section{Quark anti-quark internal energy and entropy}\label{secen1}
It has been argued that the quark anti-quark free energy,
$F_1(r,T)=U_1(r,T)-TS_1(r,T)$, contains non-trivial entropy contributions, {\em
  i.e.} $S_1=S_1(r,T)$, which, in particular, could make the analysis of
thermal modifications of the finite temperature potential complicated
\cite{Kaczmarek:2002mc,Zantow:2003ui}. We briefly discuss here the quark
anti-quark internal energy\footnote{As we calculate the internal energy and entropy using the renormalized free energies, $U_1(r,T)$ and $S_1(r,T)$ are properly fixed by construction. However, the quark anti-quark internal energy and entropy could also be seperately renormalized non-perturbatively at small distances, {\em i.e.} $U_1(r\ll1/T,T)\simeq V(r)$ and $S_1(r\ll1/T,T)\simeq0$.

It is interesting to note here that in contrast to $F_1(r,T)$ and $U_1(r,T)$, the the entropy is free from any finite renormalization at $T=0$ which, in particular, could introduce flavor dependent over-all constant contributions to the finite temperature energies.}, $U_1(r,T)=-T^2\partial(F_1/T)/\partial T$, and
entropy, $S_1(r,T)=$ \\$-\partial F_1/\partial T$.

It is important to realize from (\ref{PT}) that although the entropy
contribution at (infinite) large distances, $S_\infty(T)$, will vanish in the
perturbative high temperature limit, the quantity $TS_\infty(T)$ will increase
and dominate the difference between free and internal energy, {\em i.e.}
\begin{eqnarray}
U_\infty(T)-F_\infty(T)&=&TS_\infty(T)\;\simeq\;+{\cal O}(g^3T)\;.
\end{eqnarray} 
In the limit of zero temperature, however, the observable $TS_\infty(T)$ is
supposed to vanish as $S_\infty(T)$ is a dimension-less quantity. A qualitative
change of the temperature dependence of the observable $TS_\infty(T)$ when
going from $T=0$ to high temperatures, $T\to\infty$, is not obvious and if
present will demonstrate the phase change from confinement to deconfinement.
   
\begin{figure}[t]
  \epsfig{file=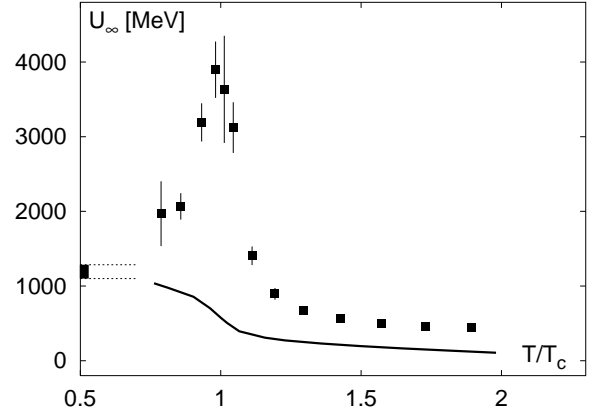,width=8.5cm}
\caption{
  The asymptotic free energy, $F_\infty$ (line), and the internal energy,
  $U_\infty$ (filled symbols), in 2-flavor QCD as function of $T/T_c$. The
  dashed band shows again $V(r\equiv\infty)\simeq1100\;-\;1285$ MeV
  \cite{Digal:2001iu,Pennanen:2000yk}.}
\label{string_breaking_paper}
\end{figure}
Our results for $F_\infty(T)$ and $U_\infty(T)$ are summarized in
Fig.~\ref{string_breaking_paper} as function of $T/T_c$. The observable,
$TS_\infty(T)=U_\infty(T)-F_\infty(T)$, can easily be deduced when comparing
the values for $U_\infty(T)$ with the values for $F_\infty(T)$.  It can clearly
be seen that $U_\infty(T)$ and $TS_\infty(T)$ exhibits a sharp peak at $T_c$.
The internal energy falls from a value of $U_\infty(T)\approx4000~\mathrm{MeV}$
at $T_c$ to about half of this value at a temperature of $T/T_c\simeq0.8$.
While the temperature dependence of $U_\infty(T)$ is quite strong in the
vicinity of the transition and markedly different from the free energy, both
quantities show a similar $T$-dependence for $1.5\;\gsim\;T/T_c\;\gsim\;2$.
This is also expected at lower temperatures ($T\;\lsim\;0.7T_c$) were
$U_\infty(T)$ and $F_\infty(T)$ are expected to smoothly approach the same
value at $T=0$, $V(r\equiv\infty)\simeq1100\;-\;1285$ MeV (from
\cite{Digal:2001iu,Pennanen:2000yk}).

\begin{figure}[t]
  \epsfig{file=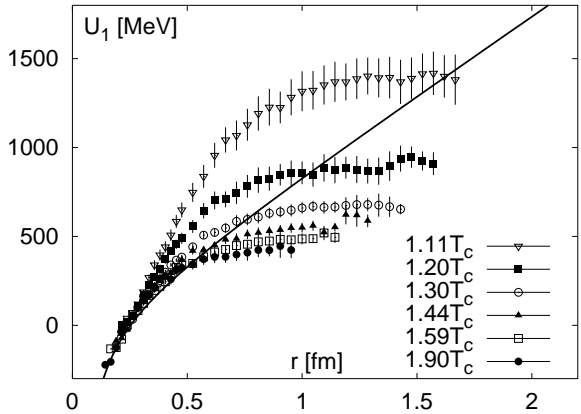,width=8.5cm}
\caption{
  The color singlet quark anti-quark internal energy above $T_c$. The solid
  line represents the $T$=0 heavy quark potential, $V(r)$.  }
\label{pots_paper}
\end{figure}

Finally we show in Fig.~\ref{pots_paper} the $r$-dependence of the internal
energies above $T_c$. Indeed $U_1(r,T)$ agrees with the zero temperature
potential at small distances while at intermediate distances $U_1(r,T)$ has a
steeper slope and clearly lies well above $F_1(r,T)$ (compare with
Fig.~\ref{plcsf}). Further details on our analysis of the finite temperature
energies, $F_1(r,T)$ and $U_1(r,T)$, and the entropy will be given
\cite{okacz05/1,okacz05/2}.

\section{Conclusions}
We have discussed thermal modifications of quark anti-quark free and internal
energies in the color singlet channel in QCD with dynamical quarks ($N_f=2$)
and compared our findings to similar studies in quenched and $3$-flavor QCD.
Although our comparison with the quenched case is in parts only on a
qualitative level, we already leads to useful information for the study of
heavy quark bound states in the plasma phase. At temperatures well above $T_c$,
{\em i.e.} $1.2\;\lsim\;T/T_c\;\lsim\;2$, no or only little differences appear
between results obtained in quenched and $2$-flavor QCD. This might suggest
that using thermal parameters extracted from energies in quenched QCD as input
for model calculations is a reasonable approximation. This also supports direct
investigations of heavy quarkonia in quenched lattice QCD
\cite{Datta:2003ww,Asakawa:2003re,Asakawa:2002xj}. We note, however, that most
of our $2$-flavor QCD results differ from quenched calculations at temperatures
in the vicinity and below the transition. These differences could make a
discussion of possible signals for the quark gluon plasma production based on
quenched QCD results complicated when temperatures close to the transition
become important.

\subsection*{Acknowledgments}
We thank the Bielefeld-Swansea collaboration for providing us their
configurations with special thanks to S. Ejiri. We would like to thank E.
Laermann and F. Karsch for many fruitful discussions. F.Z. thanks P. Petreczky
for his continuous support. This work has partly been supported by DFG under
grant FOR 339/2-1 and by BMBF under grant No.06BI102 and partly by contract
DE-AC02-98CH10886 with the U.S. Department of Energy. At an early stage of this
work F.Z. has been supported through a stipend of the DFG funded graduate
school GRK881.

\bibliographystyle{h-physrev3} \bibliography{paper}

\end{document}